\documentclass[arxiv]{SciPost}

\hypersetup{
	colorlinks,
	linkcolor={red!50!black},
	citecolor={blue!50!black},
	urlcolor={blue!80!black}
}

\usepackage{amsmath,amssymb,amsfonts,stmaryrd,wasysym,graphicx,multirow,color,textcomp,nicefrac,float,enumitem,physics}
\usepackage[bitstream-charter]{mathdesign}
\DeclareSymbolFont{usualmathcal}{OMS}{cmsy}{m}{n}
\DeclareSymbolFontAlphabet{\mathcal}{usualmathcal}

\fancypagestyle{SPstyle}{
	\fancyhf{}
	\lhead{\colorbox{scipostblue}{\bf \color{white} ~SciPost Physics }}
	\rhead{{\bf \color{scipostdeepblue} ~Submission }}
	
	\fancyfoot[C]{\textbf{\thepage}}
}

\let\linenumbers\relax      % \linenumbers 명령 제거
\begin{document}
	
	\pagestyle{SPstyle}
	
	\begin{center}{\Large \textbf{\color{scipostdeepblue}{
					Topological transition as a percolation of the Berry curvature
	}}}\end{center}
	
	\begin{center}\textbf{
			Hanbyul Kim\textsuperscript{1},
			Taewon Yuk\textsuperscript{1} and
			Sang-Jin Sin\textsuperscript{1$\star$}
	}\end{center}
	
	\begin{center}
		{\bf 1} Department of Physics, Hanyang University, Seoul 04763, Republic of Korea
		\\[\baselineskip]
		$\star$ \href{mailto:sangjin.sin@gmail.com}{\small sangjin.sin@gmail.com}
	\end{center}
	
	\section*{\color{scipostdeepblue}{Abstract}}
	\textbf{\boldmath{
			We first study the importance of the sign of the Berry curvature in the Euler characteristic of the two dimensional topological material with two band. Then we report an observation of a character of the topological transition as a percolation of the sign of the Berry curvature. The Berry curvature F has peaks at the Dirac points which enables us to divide the Brillouin zone into two regions depending on the sign of the F : one with the same sign with  a peak and the other   with the opposite sign. We observed that when the  Chern number is non-zero, the oppositely signed regions are localized, while in the case of a trivial topology, the oppositely signed regions are delocalized dominantly.   Therefore, under the topological phase transition from non-trivial to trivial, the oppositely signed region will percolates.   We checked this for several models including Haldane model, the extended Haldane model and the QWZ model. 	Our observation may serve as a novel feature of the topological phase transition.
	}}
	
	\vspace{\baselineskip}
	
	\noindent\textcolor{white!90!black}{%
		\fbox{\parbox{0.975\linewidth}{%
				\textcolor{white!40!black}{\begin{tabular}{lr}%
						\begin{minipage}{0.6\textwidth}%
							{\small Copyright attribution to authors. \newline
								This work is a submission to SciPost Physics. \newline
								License information to appear upon publication. \newline
								Publication information to appear upon publication.}
						\end{minipage} & \begin{minipage}{0.4\textwidth}
							{\small Received Date \newline Accepted Date \newline Published Date}%
						\end{minipage}
				\end{tabular}}
		}}
	}
	
	\linenumbers
	
	\vspace{10pt}
	\noindent\rule{\textwidth}{1pt}
	\tableofcontents
	\noindent\rule{\textwidth}{1pt}
	\vspace{10pt}
	
	\section{Introduction}
	Much progress has been made in the study of topological materials  in  both  theory and experiment, profoundly impacting our understanding of electronic properties and phase transitions. Initiated by the  Quantum Hall Effect\cite{QHE}, which demonstrated  capacity of topological concepts  to elucidate complex quantum behaviors, research has diversified into areas such as topological insulators \cite{QSHI,fractionalC} and topological superconductors \cite{TopologicalSuperconductivity, SatoTopologicalSC}. These studies have grounded theoretical frameworks on the topological invariant such as the Chern number, the integral of the Berry curvature over the Brillouin Zone (BZ). 

{ 		More recently, quantum geometry  played  a pivotal role  in  the study of topology \cite{Qg1}. Quantum geometry fundamentally describes how geometric and topological properties of electron wave functions influence material behavior at the quantum level \cite{Qg2,QgPolar,qgt11} . This field explores a variety of phenomena, from the influence of quantum metrics on superconducting states to the broader implications of electron band topology in designing novel materials \cite{topologyHao,Ecouplingl,qgt12, qgt13} . Specifically, quantum geometry plays crucial role in unraveling the geometric contributions to physical properties such as superconductivity and super fluidity, particularly evident in the study of super fluid weight in flat bands and similar system \cite{superfluid,superconductor}.   }

Despite the theoretical and experimental importance of the quantum metric, a comprehensive understanding of how local spatial properties, such as the percolation of Berry curvature, influence the characterization of topological phases remains elusive. As a imaginary part of the quantum metric tensor, the significance of the spatial distribution of the Berry curvature must be underscored \cite{TormaEssay, berry2}.  {This prompts the question: Is there a criterion for identifying the presence of topological order based on pattern recognition? Previous research has shown that the triviality or non-triviality of topology cannot be determined solely by the sign of the dominant peak of the Berry curvature. Here, we propose that by considering the oppositely signed regions (OSR) relative to the Berry curvature’s peak, it is possible to identify whether the system is topologically trivial or non-trivial without directly calculating the Chern number.}
Our study aims to bridge this gap by focusing on the spatial dynamics of Berry curvature and its impact on the topological properties of materials. By elucidating the localization and delocalization behaviors of Berry curvature in various topological models, we hope to provide a refined   distinction between trivial and non-trivial topological phases.

This paper explores the phenomenon of Berry curvature percolation and its impact on the topological phases of various models, including the Qi-Wu-Zhang (QWZ) model, the Haldane model and the Extended Haldane model. We specifically focus on the localization and delocalization properties of Berry curvature in relation to the Chern number, which helps to differentiate between trivial and non-trivial topological phases.
	
		\section{ Quantum Metric and Topological invariant
	}
	\label{sec:TopologySign}
	In this section, we describe about the relation of the quantum metric and the Berry curvature.

	For  an orthonormal eigenvector $|\boldsymbol{n}\rangle$ of the system with non-degenerate eigenvalues the Quantum Geometry Tensor (QGT) is
	\begin{align}
		Q_{ij}^{(n)}= \langle \partial_{i}\boldsymbol{n}| \partial_{j}\boldsymbol{n} \rangle -\langle \partial_{i}\boldsymbol{n}| \boldsymbol{n} \rangle \langle \boldsymbol{n}| \partial_{j}\boldsymbol{n} \rangle.
		\label{Eq:QGT}
	\end{align}
	The real part of the QGT  serves as the Quantum Metric   (QM), while the imaginary part of the QGT is  the Berry curvature :
	$$g_{ij}^{(n)}=\textrm {Re}Q_{ij}^{(n)}, \quad  {F}_{ij}^{(n)}= -\textrm {2Im} Q_{ij}^{(n)}.$$
	
	If we consider two band system in two dimension, the Berry curvature can be express as $ {F}_{xy}$.
	A general two band Bloch Hamiltonian reads 
	\begin{align}
		H=h_0(\boldsymbol{k})I_2+\boldsymbol{h}(\boldsymbol{k})\cdot\boldsymbol{\sigma},
	\end{align}
	where $I_2$ is the $2 \times 2$ identity matrix, $  \boldsymbol{k}=(k_x,k_y)$,  $$\boldsymbol{h}=(h_1(\boldsymbol{k}),h_2(\boldsymbol{k}),h_3(\boldsymbol{k})), \quad  \boldsymbol{\sigma}=(\sigma_1,\sigma_2,\sigma_3) ,$$  and $\sigma_i $'s  are the Pauli matrix. With $\boldsymbol{\hat{h}} = {\boldsymbol{h}}/{|\boldsymbol{h}|}$, the Berry curvature can be written as \cite{bcumath}
	\begin{align}
		{F}_{xy}
		=\frac{1}{2|\boldsymbol{h}|^3}\left[\left(\boldsymbol{h}\cdot\left(\partial_{k_x}\boldsymbol{h}\times\partial_{k_y}\boldsymbol{h}\right)\right]\right. %\notag 
		=	\frac{1}{2}\boldsymbol{\hat{h}}\cdot\left(\partial_{k_x}\boldsymbol{\hat{h}}\times\partial_{k_y}\boldsymbol{\hat{h}}\right).
		\label{eq:BerryCurvature}
	\end{align}
	In this formula, $\partial_{k_x}\boldsymbol{\hat{h}}\times\partial_{k_y}\boldsymbol{\hat{h}}$ terms describes the orientation of the surface on the Bloch sphere. 
	%This means that the Berry curvature suggests by the dot product between the vector field $\boldsymbol{\hat{h}}$ and the orientation of the surface on it. 
	The Berry curvature possesses both positive and negative values, indicating that its orientation is not well defined;   the    orientation reversal occurs along line of  $ {F}_{xy}=0$.

	To see the relation of the quantum metric and the Berry curvature, we calculate the quantum metric.
	quantum metric that can be simplified as the following:
	\begin{gather}
		g_{ij}=\frac{1}{4}\partial_{k_i}\boldsymbol{\hat{h}}\cdot\partial_{k_j}\boldsymbol{\hat{h}}.
		\label{qgt}
	\end{gather}
	With the Eq.\eqref{qgt}, the determinant of the quantum metric can be written as
	\begin{align}
		\det(g)
		=\frac{1}{16}\left[\left(\boldsymbol{\hat{h}}\cdot\left(\partial_{k_x}\boldsymbol{\hat{h}}\times\partial_{k_y}\boldsymbol{\hat{h}}\right)\right]^2\right. . \label{detg}
	\end{align}
	From this and   Eq.  (\ref{eq:BerryCurvature}),
	\begin{align}
		\det(g)=\frac{1}{4} {F}_{xy}^2,
		\label{detgF}	\end{align}
	for the two-dimensional two band system. 
	The Chern number  
	\begin{align}
		\ C =\frac{1}{2\pi} 
		\oint_{BZ}  {F}_{xy} dk_x dk_y
		\label{eq:Chernnumber}, 
	\end{align}
	is an integer as one may verify explicitly.   
	
	The Euler characteristic, $\chi$, is  a topological invariant  that can be defined by   integral of Euler class \cite{Nakahara, EulerNumberRieman}. Here we follow slightly different path, using  the Kronecker's index\cite{ EulerNumberRieman}. 
	Let $S$ be an $n$ dimensional closed Riemann manifold on which a set of $n+1$ functions    $V^i (x)$  satisfying $V^i (x)V^i (x)=1$ is defined. With this set of functions, we can consider a continuous mapping of $S$ upon the unit $n$ sphere. Define determinant $D$ with $V$ class as below:
	\begin{align}
		D =
		\begin{vmatrix}
			V^1 &... & V^{n+1}\\ \partial_{x^1} V^1 &... &  \partial_{x^1} V^{n+1}\\  \vdots & ... & \vdots \\ \partial_{x^n} V^1 &... &  \partial_{x^n} V^{n+1}
		\end{vmatrix}.
		\label{D}
	\end{align}
	Then, we can  show that  
	\begin{align}
		D^2 =\det G,  \quad  
		G_{\alpha \beta}  =  
		\partial_{x^\alpha} V \cdot \partial_{x^\beta} V , 
		\label{D2}
	\end{align}
	and define the Euler characteristic in even  dimension   by 
	\begin{align}
		\chi =\frac{2}{vol(S_n)} \int  \pm \sqrt{D^2} d^nx ,
	\end{align}
	where $vol(S_n)$ is the area of n-dimensional sphere  and  the sign is chosen to allow for overlapping of the covering \cite{ EulerNumberRieman}. 
	In our study, we consider only two dimensions in momentum space on the Bloch sphere represented by  $\hat{\boldsymbol{h}}$. 
	In two dimension case, we can express $D = 2\mathcal{F}_{xy}$ and  $G_{ij}=4g_{ij}$. 
	By using Eq. (\ref{qgt}) and Eq. (\ref{detgF}), specifically we can simplify $D^2$ as 
	\begin{align}
		D^2=\det G= 16 \det g  = 4 \mathcal{F}_{xy}^2. 
		\label{DFrelation}
	\end{align}
	
	From Eq .(\ref{DFrelation}), the Euler characteristic number can be expressed as follows :
	
	\begin{align}
		\chi= \frac{2}{\pi} \int_{S} \pm \sqrt{\det g} dk_x dk_y = \frac{1}{\pi} \int_{S} \mathcal{F}_{xy} dk_x dk_y.
	\end{align}
	
	Therefore, $\chi$ can be simply   represented by: 
	\begin{align}
		\chi =2C.
	\end{align}
	The important  point to notice is  that the existence of overlapping regions in the mapping of the Brillouin zone (BZ) to the Bloch sphere causes an orientation reversal. Specifically, there is an induced orientation on the BZ due to the pullback of the mapping from momentum space to the manifold of quantum states, denoted as
	$${\bf \widehat h}: BZ\to \hbox{Bloch  sphere} ,$$
	which plays the role of the Gauss map in the theory of the surface. 
	It turns out that the sign of the Berry curvature determines this orientation \cite{Riemann}. Thus, due to the orientation reversal across the BZ along the line where $  {F}_{xy}=0$, the orientation sign must be considered carefully—an aspect often neglected in many preceding studies. Therefore, the sign of the Berry curvature serves as a crucial indicator for observing the quantities associated with topology.
	
		\section{The percolation of the Berry curvature }
	In the last section, we observed that   the sign of the Berry curvature serves as crucial means  in the mechanism of  the  topological observable to work. 	In this section, we explores the crucial phenomenon of Berry curvature percolation and its substantial role in determining topological phases. A fundamental aspect of this discussion is understanding how localization and delocalization of sign of Berry curvature influence topological properties.  In two-dimensional BZ, there is periodicity along $k_x$ and $k_y$ directions so we can mapping Berry curvature on torus. Initially, We define the terminology "localization" of the specific regions  having certain sign of Berry curvature as implying that the region is discontinuous along any trajectory initiated from a specific point within it, thus preventing a return to the starting point.  {On torus, "localization" means that there are contractable loops on a torus on certain signed regions. For example, FIG. 3 provides this localization with clusters. Conversely, "delocalization" refers to the continuous property of a region, allowing trajectories to return to their starting point.     When we map the specific signed region of Berry curvature, there are non-contractible loops on a torus.

	\subsection{Percolation of the Berry curvature}
	Berry curvature tends to be peaked  at specific locations where energy bands exhibit near crossings or touching. 
	To see this notice that in the 2-band system, the Berry curvature associated with + band can be expressed as
	\begin{align}
		{F}_{xy}(\boldsymbol{k})
		=& i\frac{1}{(\epsilon_+-\epsilon_{-})^2}[  \langle \boldsymbol{n_+}| \partial_{k_x} H|\boldsymbol{n_-} \rangle \langle \boldsymbol{n_-}| \partial_{k_y} H|\boldsymbol{n_+} \rangle  \notag
		\\   &\quad-\langle\boldsymbol{n_+}| \partial_{k_y} H|\boldsymbol{n_-} \rangle \langle \boldsymbol{n_-}| \partial_{k_x} H|\boldsymbol{n_+} \rangle  ], \\
		=& -\frac{2\,\text{Im}\left(\langle \boldsymbol{n_+}|\partial_{k_x} H|\boldsymbol{n_-} \rangle \langle \boldsymbol{n_-}| \partial_{k_y} H|\boldsymbol{n_+} \rangle\right)}{(\epsilon_+-\epsilon_{-})^2},
		\label{BerrycurvatureEq}
	\end{align}
	where $\epsilon_\pm$ is eigenvalue and  $| \boldsymbol{n}_\pm \rangle$ is the eigenvector for each   band.
	When there is a degeneracy of eigenvalues at a band touching point, the value of $(\epsilon_+-\epsilon_{-})^2$ approaches zero, making the peak in the Berry curvature at the near touching point. 	
	
	Now let's concentrate on the sign of the $F_{xy}$. The sign flip of  $F_{xy}$  obviously changes the distribution of the Peak Signed Regions (PSR) of the Berry curvature as well as that of the Oppositely Signed Regions(OSR). We defined PSR with the dominant peak's sign; the peaks on band touching point in topological transition
	From Fig. \ref{fig:HaldaneExample}, the PSR are depicted in red, and the OSR in blue. When the Chern number is non-trivial, the OSR is localized, resembling islands, while the PSR is predominantly delocalized. Conversely, in cases where the Chern number is trivial, the OSR is extensively delocalized, akin to a river. This pattern suggests that the topological phase transition is accompanied by the percolation of the OSR. We consistently verify this intriguing phenomenon across various models, confirming its general applicability in describing topological transitions. Furthermore, as discussed in Section \ref{sec:TopologySign}, the distribution of the Berry curvature can be equated to the distribution of the metric tensor within the framework of the QGT, further elucidating the geometric underpinnings of these topological phenomena.

	\begin{figure}[h]
		\includegraphics[width=1 \linewidth]{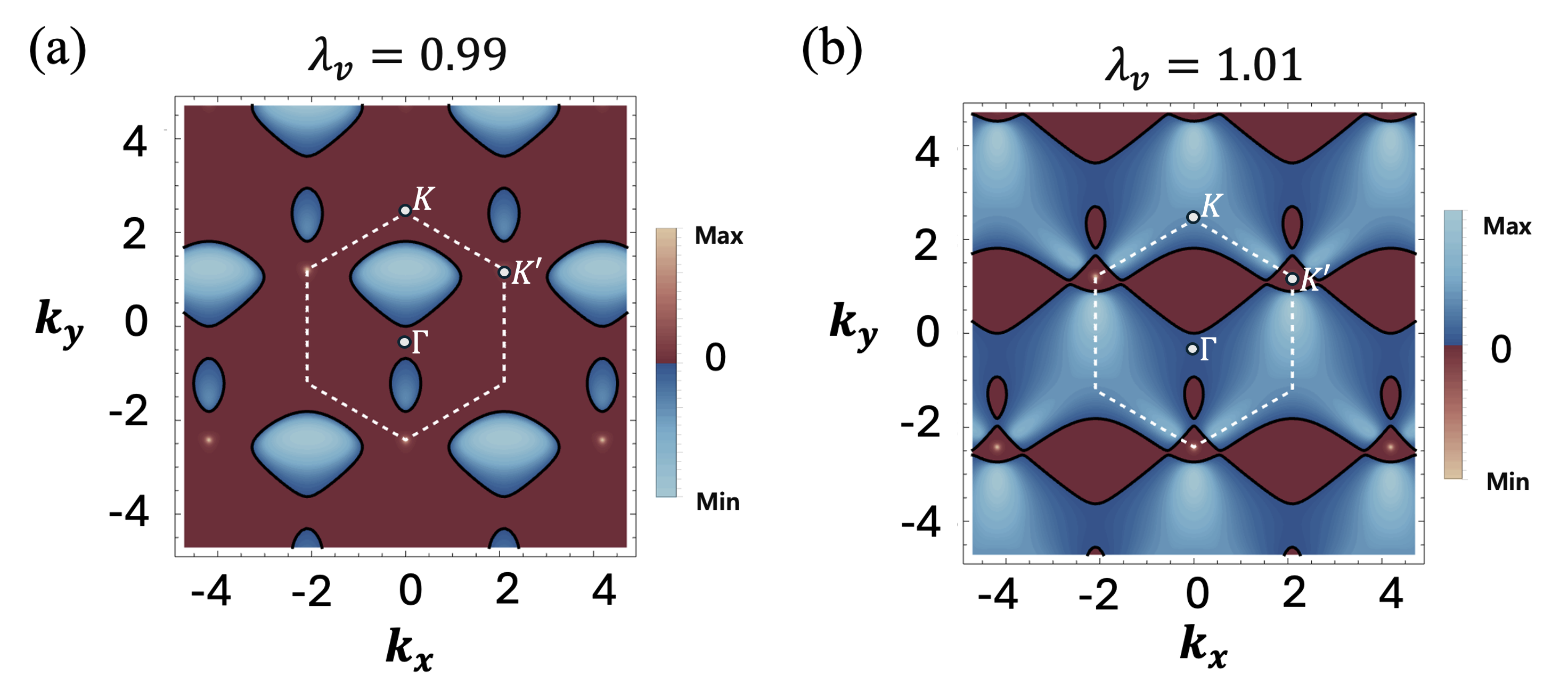}
		
		\caption{ The Berry curvature of the Haldane model.
			Red regions indicate the PSR and the blue regions describe the OSR.  Dashed line represents BZ. In (a), the Chern number is non trivial thus the OSR is localized. Conversely, in the figure (b), the OSR is delocalized when the Chern number is trivial. 	}
		\label{fig:HaldaneExample}
	\end{figure}

	\subsection{Behavior of the Berry curvature}	
	
	In this section, we elucidates the reasons for these observations, focusing on how variations in the Berry curvature's distribution directly relate to changes in the Chern number.
	
	The primary factor influencing changes in the distribution of the Berry curvature is band inversion.
	The band inversion occurs when the energy levels of orbitals, typically of differing parities, interchange at specific high-symmetry points within the BZ through a process where the band gap initially narrows, closing, and then reopens. After the band gap is reopened, orbitals in the lower valence band and upper band conduction band  around the gap closing point are inverted with respect to their trivial phase \cite{BI2}. 
	\begin{figure}[h]
		\includegraphics[width=1 \linewidth]{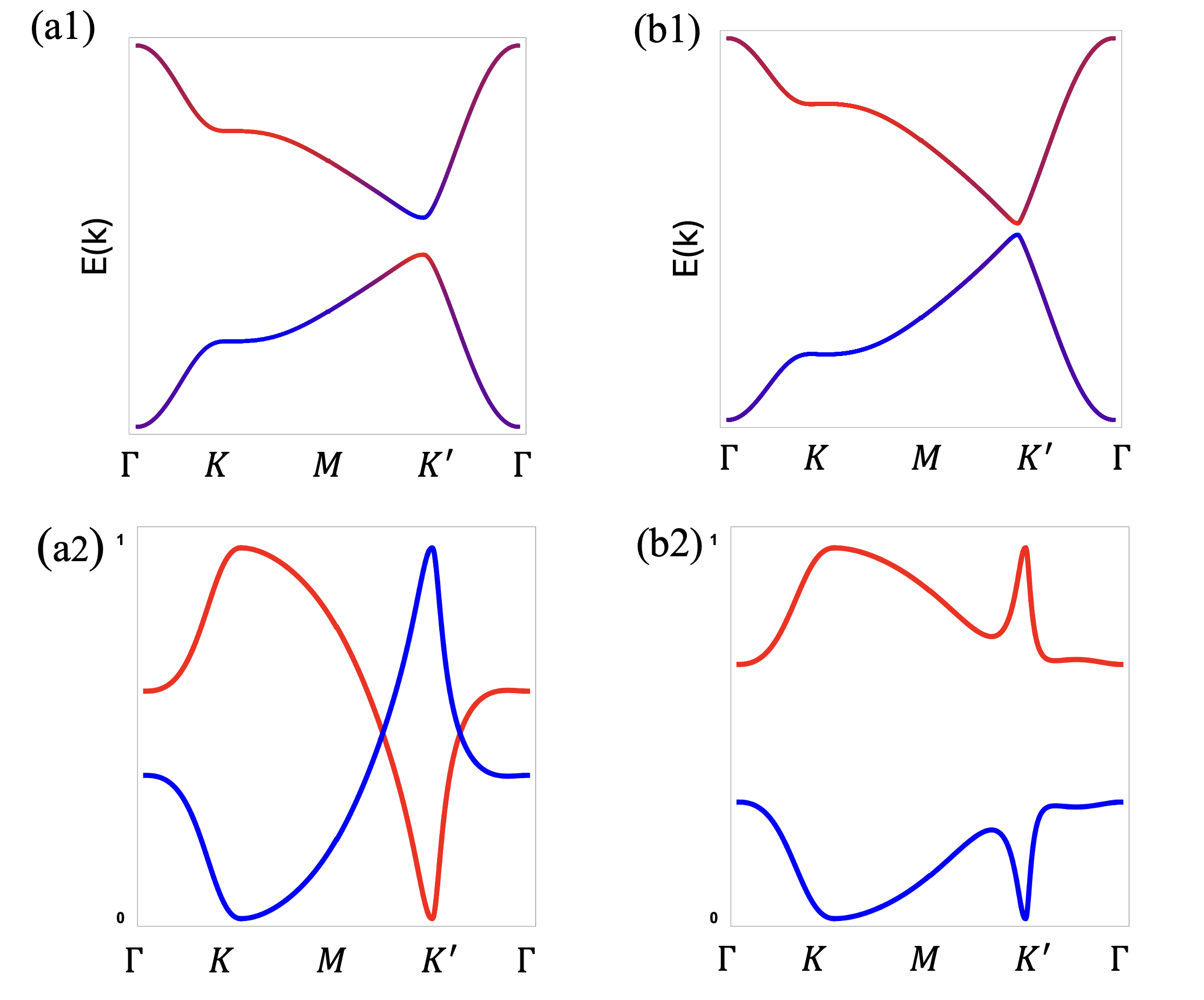}
		
		\caption{
			(a1), (b1): Illustration of the band structure of the Haldane model when topological phase is non-trivial and trivial  correspondingly. The color represents magnitude of the components of the eigenvector $\ket{\psi} = (\alpha,\beta)^\mathrm{T}$, where $\ket{\psi} = (1,0)^\mathrm{T}$ appears to be red, and $\ket{\psi} = (0,1)^\mathrm{T}$ appears to be blue. The color shift can be observed in (a1) due to the band inversion near the $K'$-point in the topologically non-trivial phase, which is not manifested in (b1). (a2) and (b2) show the magnitude of the first component of the eigenvector ($|\alpha|)$ in (a1) and (b1) respectively. Red indicates the upper band, while blue describes the lower band.		}
		\label{fig:BandInversion}
	\end{figure}
	Near the band touching point, there is a distinct region where eigenvectors from a different band are mixed (See Fig. \ref{fig:BandInversion} ). 
	Such band mixing regions  always  exists around the band touching point. 
	Consequently, in the vicinity of band touching point, the Berry curvature is predominantly constituted by eigenvectors from the other band, contrasting with other regions where the Berry curvature is composed of eigenvectors from the original band.  It  signifies that there is a eigenvector jumping near the peak. 
	
	This phenomena  in eigenvector configuration is crucial in understanding underlying nature of the distribution of the Berry curvature  under the topological phase transition: due to the mixing of eigenvectors, the sign of the Berry curvature changes near the band-touching point. 
	In brief, the band inversion modifies localization property of the Berry curvature.

	Furthermore, according to Eq. \ref{BerrycurvatureEq}, when a sufficient band gap exists, the Berry curvature values within the OSR are significantly lower than those at the peaks. Therefore, the distribution of the OSR plays a crucial role in quantitatively compensating for the peak's value, directly affecting whether the Chern number is trivial or non trivial.
	Specifically, when the Chern number is non-trivial, the OSR are localized, leading to minimal compensation for the peak's value, thereby ensuring that the Chern number remains non-zero. Conversely, when the Chern number is trivial, the flatness of the OSR necessitates their dominant distribution across the Brillouin Zone to sufficiently compensate for the peaks’ values, consequently resulting in their delocalization.
	Overall, the analysis of the Berry curvature distribution serves as a critical indicator of topological characteristics, determining whether the topological phase is trivial or non-trivial.
	
	\subsection{Percolation rate}
	Percolation theory, a critical concept in statistical mechanics, studies the behavior of connected clusters in a random medium. It provides insights into phase transitions by analyzing the probability and geometry of clusters as they form and spread through a system \cite{percolation}.
	Percolation refers to the phenomenon where localized clusters spread and merge, during a phase transition. 
	
	In this context, the delocalization of the OSR can be quantified by considering the OSR as clusters in the percolation theory. We define $S_i$ as the size of individual $i$-$th$ OSR clusters, considering periodicity as shown in Fig. \ref{fig:Percolation rate}. Then, the percolation rate of the Berry curvature can be written as follows,
	\begin{align}
		P=\frac{S_{max}}{N},
		\label{percolationRate}
	\end{align}
	where $N$ is the size of the BZ, and $S_{max}$ is the size of the largest cluster. 	
	When the Chern number is non trivial, the OSR remains localized, forming several small clusters, resulting in a low percolation rate of under 0.5. Conversely, when the Chern number is trivial, the OSR delocalizes into one connected cluster, leading to a high percolation rate of over 0.5. 
	
	The intriguing point is that delocalization of the OSR occurs always dominantly in the BZ when the Chern number is trivial. If there are narrow delocalized river of the OSR, and the dominant localization of the PSR, then there sholud exist non zero Chern number. But the important thing is that case never happened : if there is delocalization of the OSR, the OSR occupy the BZ dominantly, thus Chern number of that system is always zero with a delocalization of the OSR.
	\begin{figure}[H]
		\centering
		\includegraphics[width =  1 \linewidth]{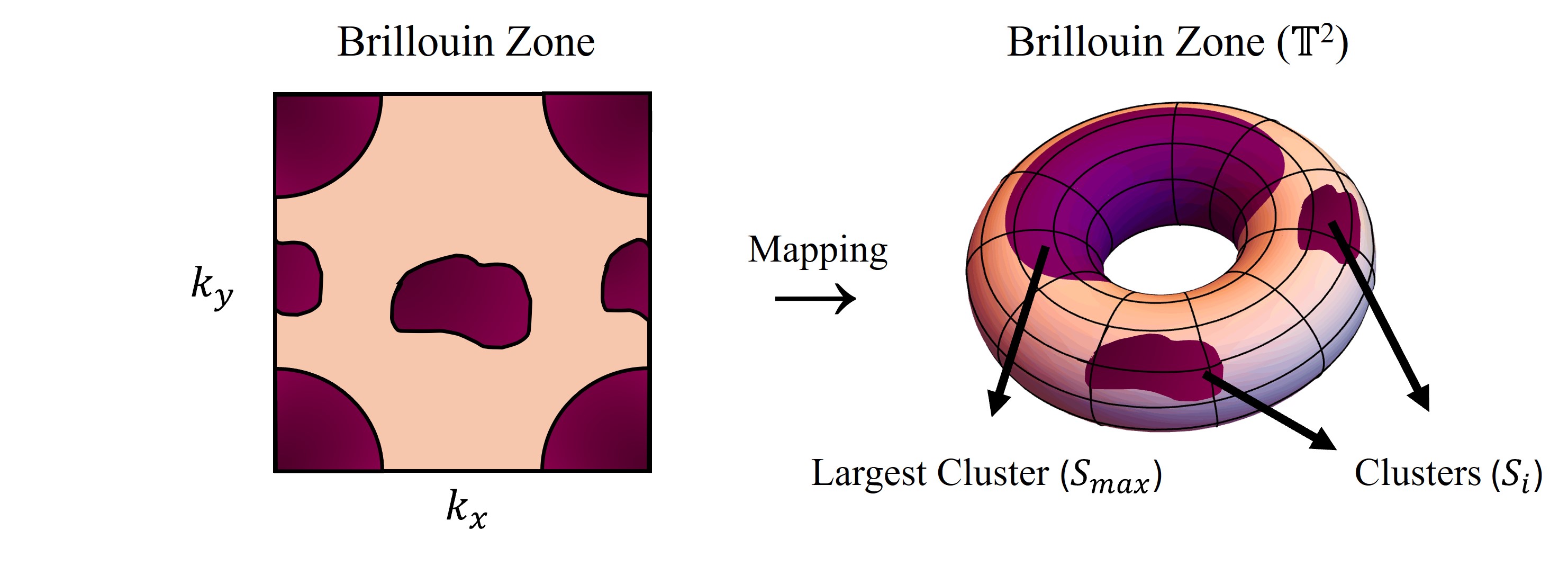}
		\caption{Percolation rate schematic diagram. Mapping the Berry curvature on the torus. Purple regions are OSR and the other regions indicate PSR.}
		\label{fig:Percolation rate}
	\end{figure}
	
	\section{Application}
	We now demonstrate how the percolation of the Berry curvature appears when the topological phase is trivial or non-trivial in three different 2D models. These models' topological invariants are defined by the Chern class. With those models, we provide contour figures of Berry curvature ; the red regions always represent PSR while the blue regions represent OSR.

	\subsection{Qi-Wu-Zhang model }
	The Qi-Wu-Zhang (QWZ) model is a two-dimensional lattice model that describes the quantum anomalous Hall effect, serving as a fundamental framework for understanding topological phases in condensed matter physics\cite{QWZorigin,Qwz1}. The Bloch Hamiltonian of the QWZ model is given by 
	\begin{align}
		h_{\text{QWZ}}= A\sin k_x \sigma_1 +A\sin k_y \sigma_2 + (\cos k_x+\cos k_y+\Delta) \sigma_3 . 
	\end{align}
	We can directly compute the Berry phase with the Hamiltonian above, as in Eq (\ref{eq:BerryCurvature}).
	
	In the QWZ model, the Chern number depends on the ratio of $A$ to $\Delta$ as follows.
	
	\begin{equation}
		\begin{cases}
			C = 1 & \text{for} \quad 0<\frac{\Delta}{A}<2 \\
			C = -1 & \text{for} \quad -2<\frac{\Delta}{A}<0 \\
			C = 0 &  \text{for} \quad \frac{\Delta}{A}>2\quad \text{or} \quad \frac{\Delta}{A}< -2 \\
		\end{cases} \notag
	\end{equation}
	
		\begin{figure*}[t!]
		\centering
		\includegraphics[width =  1 \linewidth]{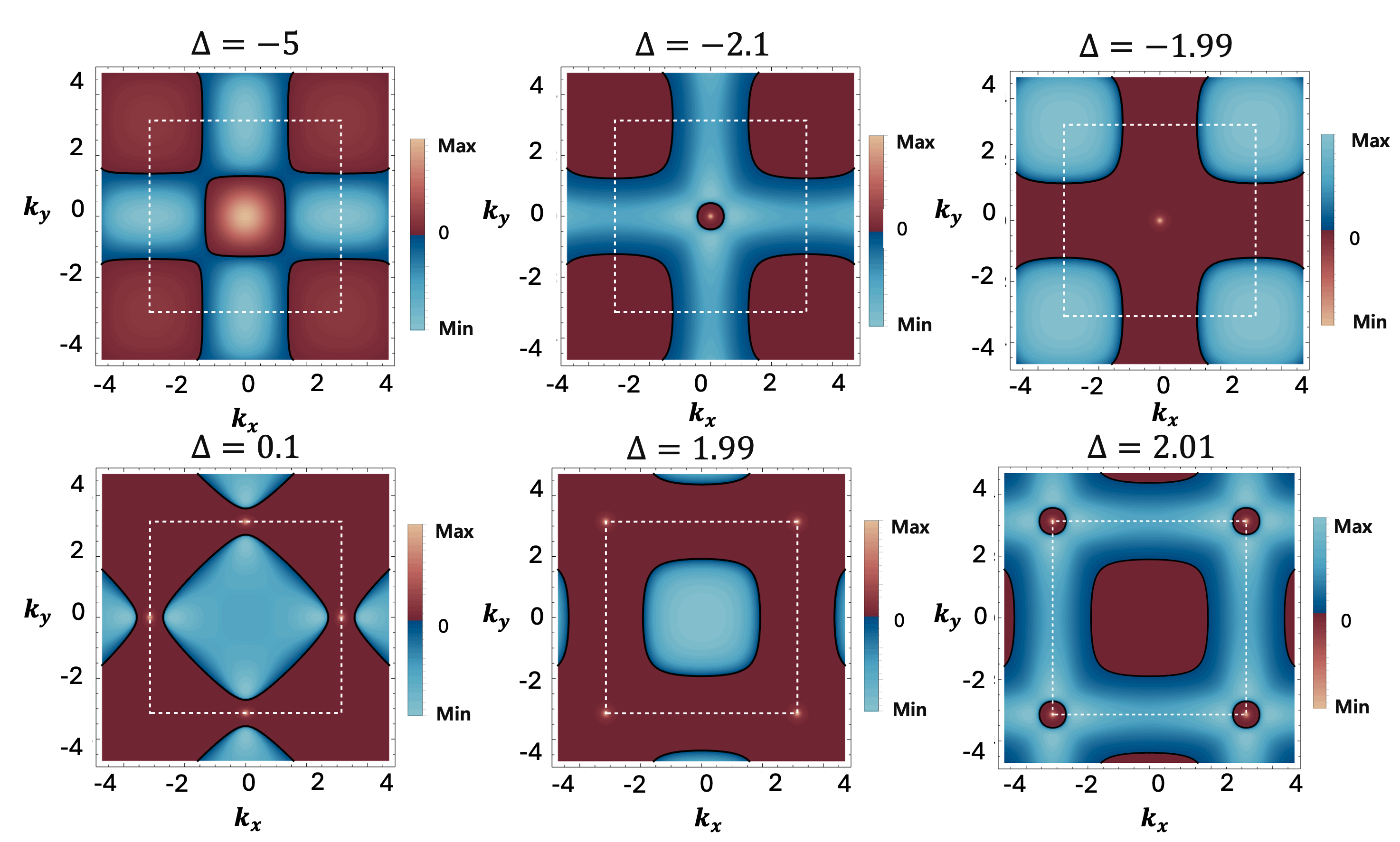}
		\caption{Berry curvature of the QWZ model when $A=1$: The red region is the PSR and the blue region is the OSR. Dashed line represents the equivalent BZ of the QWZ model. When $\Delta<-2$ or $\Delta>2$, the Chern number is trivial, so the blue area is delocalized. In the other case, when $-2<\Delta<2$, the blue region is localized while delocalization of the red regions occurs.}
		\label{fig:QWZberrycurvature}
	\end{figure*}

	In the Fig. \ref{fig:QWZberrycurvature}, dashed lines represent the BZ. The red color denotes the PSR, while blue indicate the OSR. When the Chern number is non-trivial, the PSR are predominantly delocalized while the OSR are localized like a island. $\Delta=0$ is also a topological phase transition point from -1 to 1. However, since the Chern number of both sides are non trivial, the localization of the the OSR remains. Conversely, in the trivial Chern number case, the OSR are delocalized. 
	
	Furthermore, we calculate the percolation rate with Eq (\ref{percolationRate}). In the Fig. \ref{fig:QWZPR}, the red line denotes $P=0.5$, the blue line is the percolation rate of the OSR in the QWZ model. When the Chern number is zero, the percolation rate is higher than 0.5 which means that the OSR are delocalized dominantly in the BZ. In the other case, when the Chern number is non zero, the percolation rate is lower than 0.5. 
	Consequently, observing the distribution of the Berry curvature's each signed region serves as the topological feature that determines the topological phase is trivial or non trivial.

	\begin{figure}[t]
		\includegraphics[width =  0.8 \linewidth]{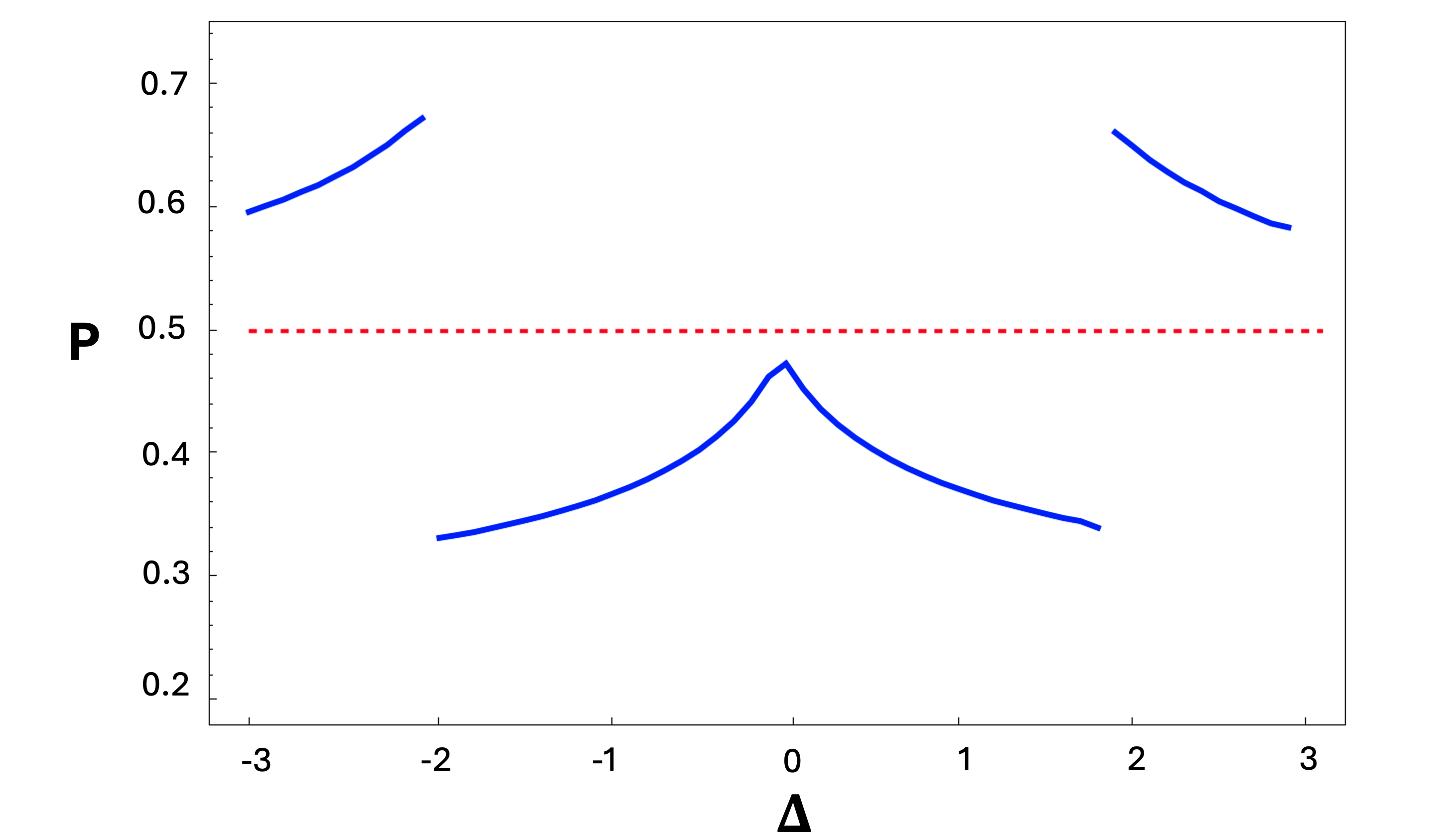}
		\caption{Percolation rate of the OSR in the QWZ model. The red line denotes $P = 0.5$. When $\Delta<-2$ or $\Delta>2$, the Chern number is trivial, so the  percolation rate is higher that the 0.5. In the other case, when $-2<\Delta<2$, when Chern number is non trivial, the percolation rate is lower than the 0.5.}
		\label{fig:QWZPR}
	\end{figure}

	\subsection{Haldane model }
	We also verify the percolation property of the Berry curvature in the Haldane model\cite{HaldaneOrigin}. 
	Haldane Model's tight binding Hamiltonian is given by

	\begin{align}
		H_{\text{H}}=t_1\sum_{\langle ij \rangle}c_i^{\dagger}c_j+t_2\sum_{\langle\langle ij \rangle\rangle}e^{-i\nu_{ij}\phi}c_i^{\dagger}c_j+\lambda_{v}\sum_{i}\epsilon_ic_i^{\dagger}c_i,
	\end{align}
	where  $\langle ij \rangle$ denotes the summation over nearest-neighbor (NN) hopping and ${\langle\langle ij \rangle\rangle}$ represents the sum over next-nearest-neighbor (NNN) interactions. $t_1$ is the hopping constant which determines NN hopping. The inclusion of a second term accounts for next-nearest-neighbor (NNN) hopping, influenced by both the internal magnetic flux $\phi$ and the hopping parameter $t_2$. Furthermore, the phase factor associated with this term depends on the direction of the NNN hopping:
	$$\nu_{ij}=sgn(\hat{d}_i\times\hat{d}_j)_z=\pm1, \quad (i,j)\in\{1,2\}$$ where $\hat{d}_{1,2},$ is the vectors along the directions of NNN hoppings. The last term is on-site energy, which depends on the sublattices:  $\epsilon_i=1$ and $\epsilon_i=-1$ for each site. 
	Here, we define a primitive lattice vector as
	\begin{align}
		\vec{\boldsymbol{d}}_1=\frac{\sqrt{3}}{2}a(\sqrt{3},1), \quad \vec{\boldsymbol{d}}_2=\frac{\sqrt{3}}{2}a(-\sqrt{3},1),
	\end{align}
	where $a$ is a lattice spacing. 
	Using the Pauli matrix, we can express the Bloch Hamiltonian by the following way:
	\begin{align}
		&h_{\text{H}}(\boldsymbol{k})=\,2t_2\cos(\phi)\left(2\cos(\tilde{k}_x)\cos(\tilde{k}_y)+\cos(2\tilde{k}_y)\right)I_2\nonumber \\ &  -t\left(1+2\cos(\tilde{k}_x)\cos(\tilde{k}_y)\right)\sigma_1 \nonumber-2t\sin(\tilde{k}_x)\cos(\tilde{k}_y)\sigma_2
		\\&+ \left(\lambda_{v}+2t_2\sin(\phi)\left(2\cos(\tilde{k}_x)\sin(\tilde{k}_y)-\sin(2\tilde{k}_y)\right)\right)\sigma_3. \label{eq:haldanetb}
	\end{align}
	where $\tilde{k}_x=3ak_x/2$, $\tilde{k}_y=\sqrt{3}ak_y/2$. \\ 
	\\
	
		\begin{figure}[H]
		\includegraphics[width = 0.8 \linewidth]{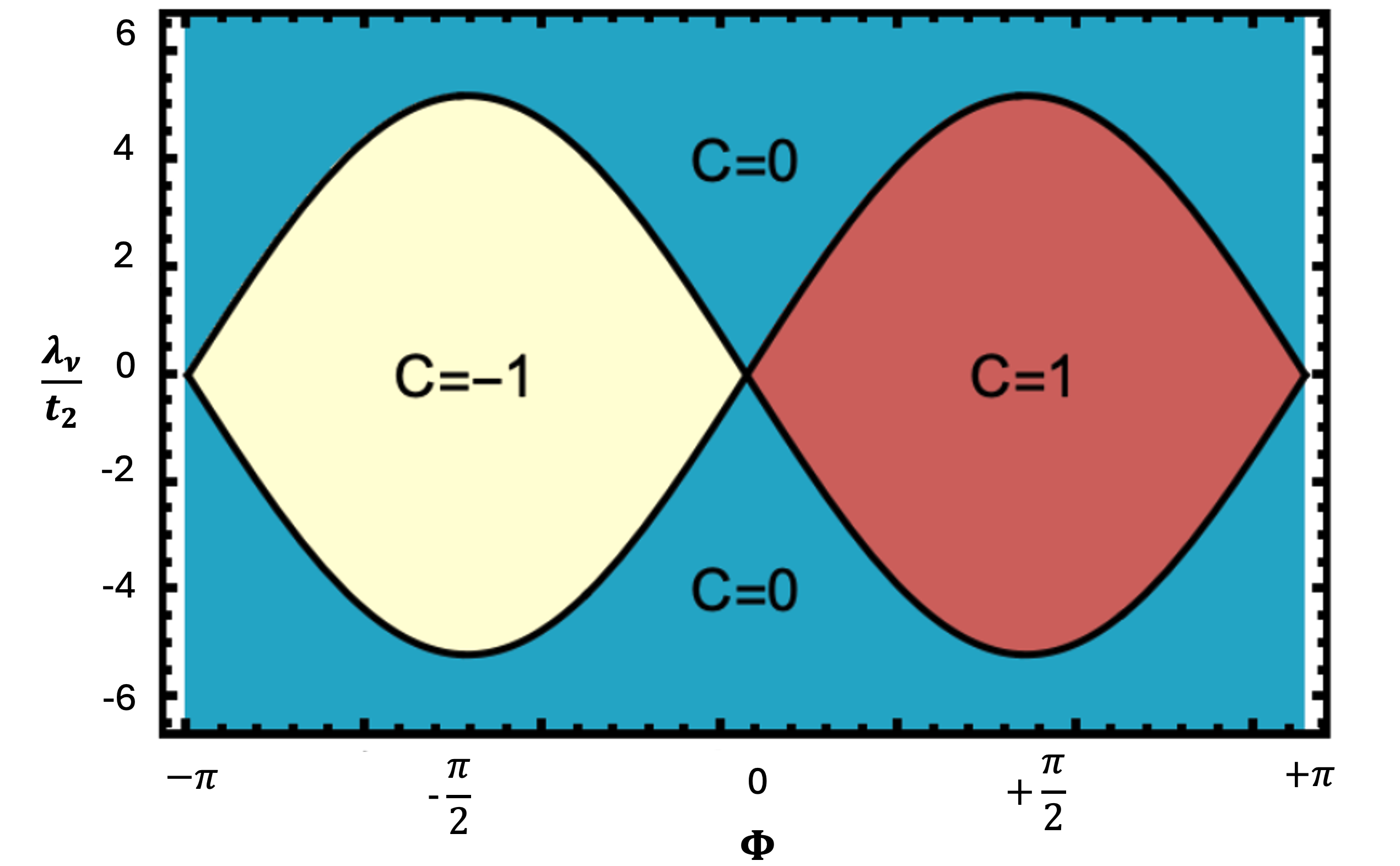}
		\caption{Phase diagram of the Haldane model \\  }
		\label{fig:phasediagramHaldane}
		\end{figure}

	Utilizing Eq (\ref{eq:Chernnumber}), the Chern number of the Haldane model is determined by the ratio of the three independent parameters : $t2, \phi,$ and $\lambda_v$. A detailed representation of these dependencies is illustrated in the phase diagram, presented in Fig. \ref{fig:phasediagramHaldane}.

	In the Fig. \ref{fig:HaldaneBerrycurvature}, dashed lines represent the BZ. We fix $\phi=\pi/2$ while varying the ratio of the $\lambda_v$ to ${t_2}$ and investigate the delocalization properties of the PSR and the OSR. Similar to the observations in the QWZ model, in the Fig. \ref{fig:HaldaneBerrycurvature}, the red areas represent the PSR, whereas the blue areas is the OSR. As demonstrated in Fig. \ref{fig:HaldaneBerrycurvature}, in the non-trivial phase where $\lambda_v < 1$ , delocalization of the PSR,  become extensive, akin to a river while the OSR are localized. Conversely, in the trivial phase, when the Chern number is zero and $\lambda_v > 1$, the OSR become  delocalized with a topological transition which represent the percolation of the OSR. 
	
	Furthermore, the percolation rate is observed as Fig. \ref{fig:HaldanePR}. When the $-1< \lambda_v < 1$, the OSR are localized in specific regions, thus the percolation rate is lower than 0.5. On the other hand, when the $-1> \lambda_v \ or \ 1< \lambda_v $ , the delocalization of the OSR occurs dominantly in the BZ, so the percolation rate is higher than 0.5.
	Consquently, the percolation of the OSR occurs with the topological transition.
	
		\begin{figure*}[t!]
		\includegraphics[width =  1\linewidth]{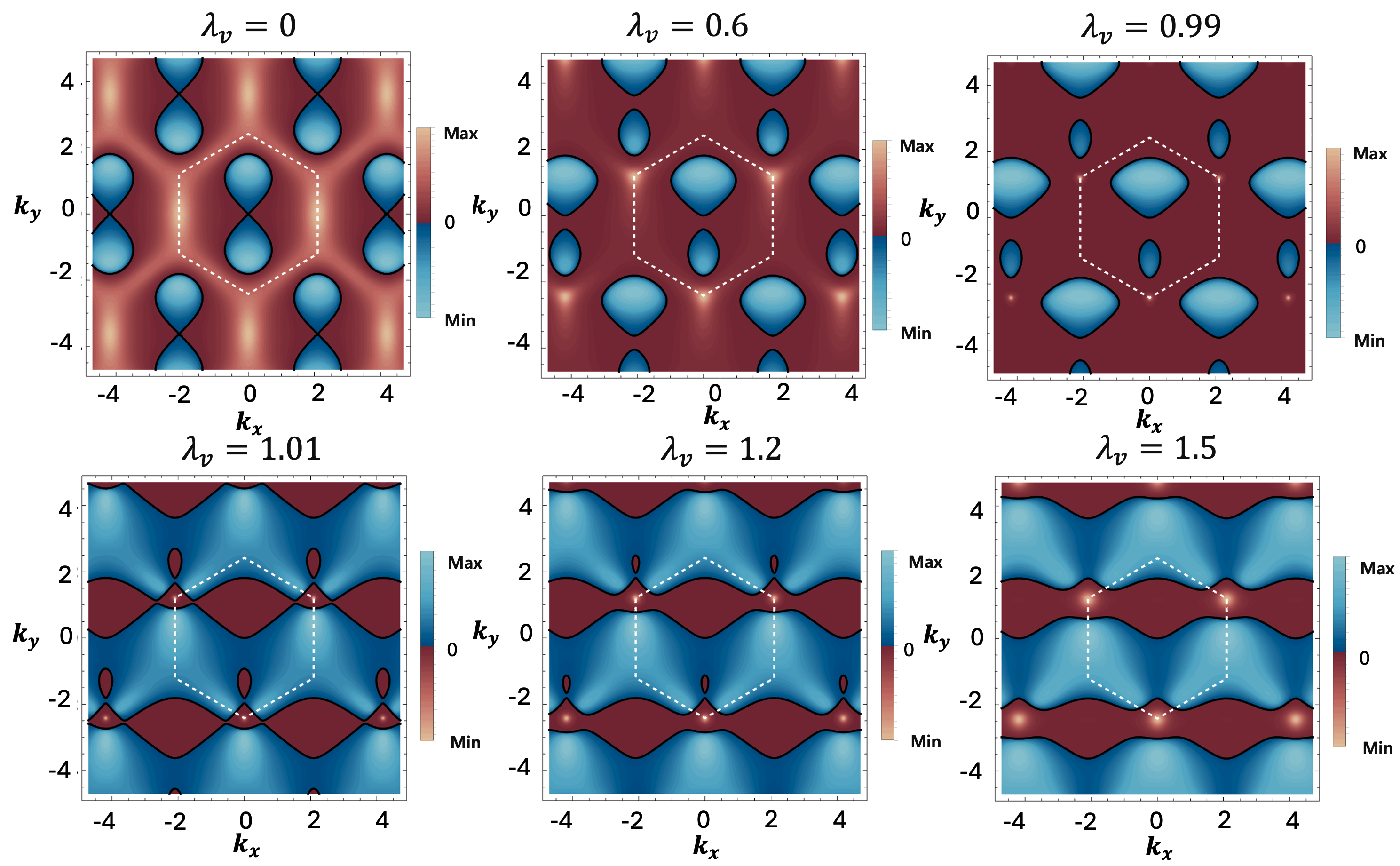}
		\caption{Berry curvature of the Haldane model where $t_2=1/3\sqrt{3}$ and $\phi = \pi/2$ : Dashed line represents the equivalent BZ of the Haldane model. Analogous to the previous figure, the red region is peak signed region and the blue region is the oppositely signed region. When the $\lambda_v < 1$, the topological phase is non trivial, therefore the percolation of the red region occurs with a localization of blue region. On the other hand, the topological phase is trivial, the blue region is delocalized. }
		\label{fig:HaldaneBerrycurvature}
	\end{figure*}
	
		\begin{figure}[H]
		\centering
		\includegraphics[width =  0.8 \linewidth]{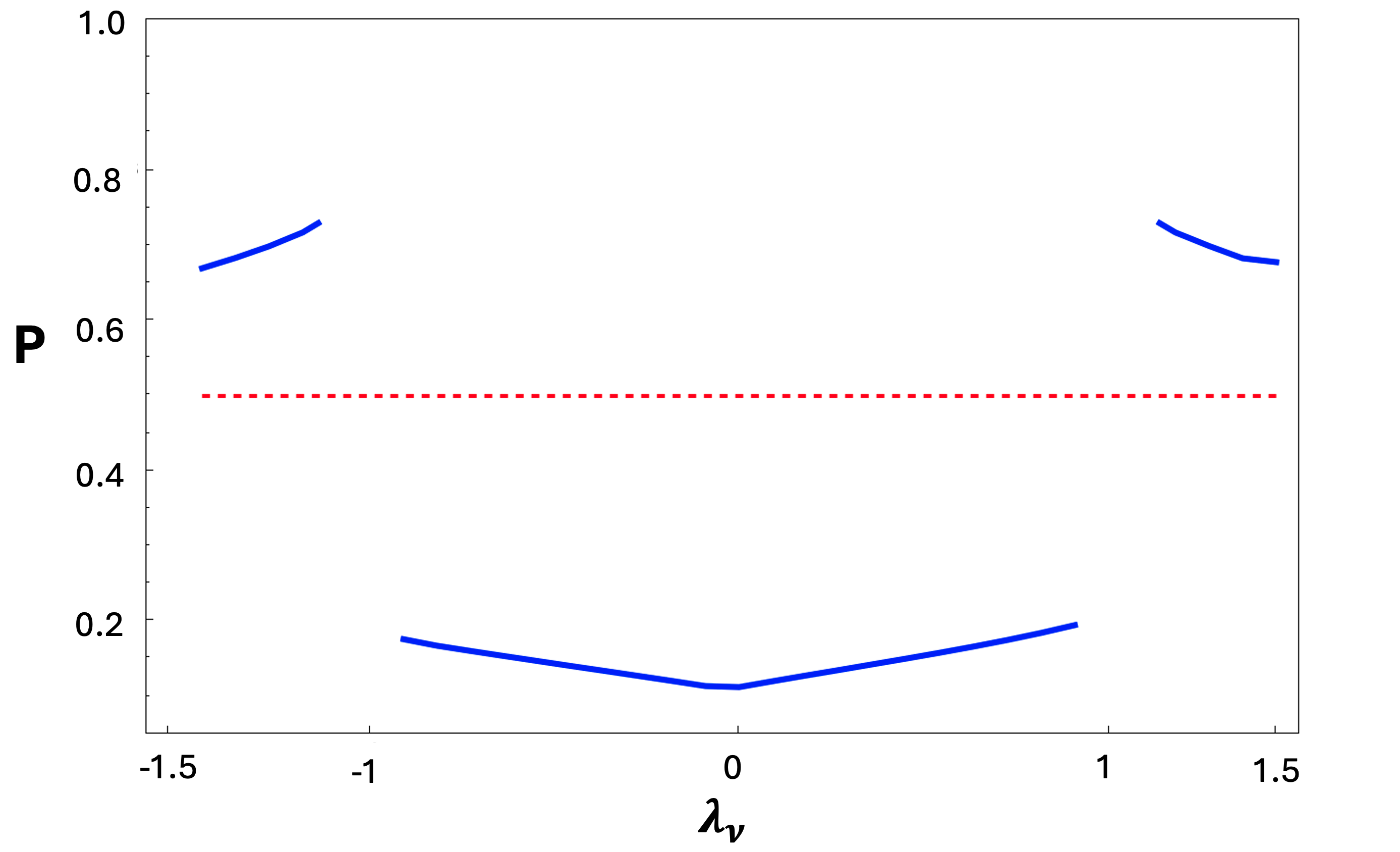}
		\caption{Percolation rate of the OSR in the Haldane model. The red line denotes $P = 0.5$. When $-1>\lambda_v$ or $ \lambda> 1$, the Chern number is trivial, the percolation rate is higher than 0.5 which means that delocalization occurs dominantly in the BZ. On the other hand, the Chern number is non trivial, the percolation rate is lower than the 0.5.}
		\label{fig:HaldanePR}
	\end{figure}

	\subsection{Extended Haldane Model}
	Extended Haldane model has next to next nearest neighbor (N3 neighbor) hopping in addition to the Haldane model \cite{N3Haldane}. 
	The Extended Haldane Model's tight binding Hamiltonian is given by
	\begin{align}
		H_{\text{EH}}= & t_1\sum_{\langle ij \rangle}c_i^{\dagger}c_j+t_2\sum_{\langle\langle ij \rangle\rangle}e^{-i\nu_{ij}\phi}c_i^{\dagger}c_j \notag
		\\&+t_3\sum_{\langle\langle\langle ij \rangle\rangle\rangle}c_i^{\dagger}c_j+\lambda_{v}\sum_{i}\epsilon_ic_i^{\dagger}c_i.
	\end{align}
	In addition to the Haldane model,  $\langle \langle \langle ij \rangle\rangle\rangle$ means the summation over N3 neighbor hopping and $t_3$ is the 
	
	Under the Fourier, the tight binding Hamiltonian can be written in Bloch Hamiltonian, 
	
	\iffalse
	\begin{align}
		h_0 = 2 t_2 \cos(\phi)[\cos{(\boldsymbol{k}\cdot\boldsymbol{a_1})} +\cos{(\boldsymbol{k}\cdot\boldsymbol{a_2})} +\cos{(\boldsymbol{k}\cdot(\boldsymbol{a_1}-\boldsymbol{a_2}))}],
	\end{align}
	\begin{align}
		h_1 = & t_1 [\cos{(\boldsymbol{k}\cdot\boldsymbol{a_1})} +\cos{(\boldsymbol{k}\cdot\boldsymbol{a_2})}] \notag \\&+t_3[\cos{(\boldsymbol{k}\cdot(\boldsymbol{a_1}+\boldsymbol{a_2}))+2\cos(\boldsymbol{k}\cdot(\boldsymbol{a_1}-\boldsymbol{a_2}))}],
	\end{align}
	\begin{align}
		h_3 = \lambda_v + M_H
	\end{align}
	\begin{align}
		M_H = 2 t_2 \sin(\phi)[\sin(\boldsymbol{k}\cdot\boldsymbol{a_2})-\sin(\boldsymbol{k}\cdot\boldsymbol{a_1})+\sin (\boldsymbol{k}\cdot(\boldsymbol{a_1}-\boldsymbol{a_2}))]
	\end{align}
	\fi
	
	\begin{align}
		h_0 = &\; 2 t_2 \cos(\phi) \big[\cos(\boldsymbol{k} \cdot \boldsymbol{d}_1) + \cos(\boldsymbol{k} \cdot \boldsymbol{d}_2) \\ &\ + \cos(\boldsymbol{k} \cdot (\boldsymbol{d}_1 - \boldsymbol{d}_2))\big], \notag \\
		h_1 = &\; t_1 \big[\cos(\boldsymbol{k} \cdot \boldsymbol{d}_1) + \cos(\boldsymbol{k} \cdot \boldsymbol{d}_2)\big] \notag \\
		&\; + t_3 \big[\cos(\boldsymbol{k} \cdot (\boldsymbol{d}_1 + \boldsymbol{d}_2)) + 2 \cos(\boldsymbol{k} \cdot (\boldsymbol{d}_1 - \boldsymbol{d}_2))\big], \notag \\
		h_3 = &\; \lambda_v + M_H,
	\end{align}
	where $M_H$ is the Haldane mass $M_H = 2 t_2 \sin(\phi) [\sin(\boldsymbol{k} \cdot \boldsymbol{d}_2) - \sin(\boldsymbol{k} \cdot \boldsymbol{d}_1) + \sin(\boldsymbol{k} \cdot (\boldsymbol{d}_1 - \boldsymbol{d}_2))]$ and $\boldsymbol{d_1}=a/2(\sqrt{3},3)$ , $\boldsymbol{d_2}=a/2(-\sqrt{3},3)$ like in the Haldane model. From Fig. \ref{fig:N3Haldanephasediagram}, we can see the the Chern number phase diagram for the Extended Haldane model.
	
	\begin{figure}[H]
	\centering
	\includegraphics[width =  0.6\linewidth]{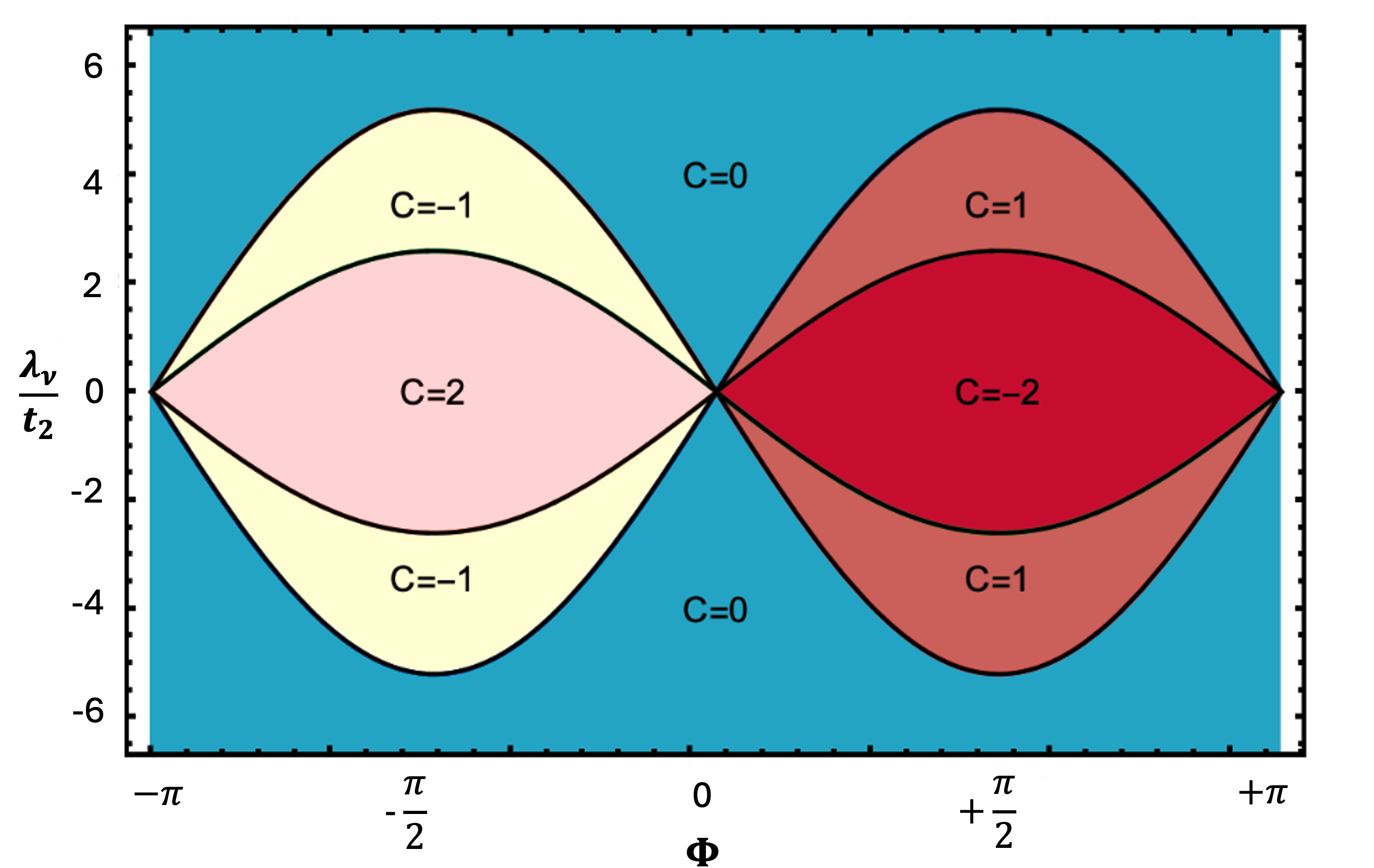}
	\caption{Phase diagram of the Extended Haldane model \\  when $t_1=1 , t_3=0.35$.}
	\label{fig:N3Haldanephasediagram}
	\end{figure}

	Similar to the Haldane model, we fixed the $\phi=-\pi/2$ and varied the ratio ${\lambda_v}/{t_2} $ continuously. When the Chern number is trivial, which occurs when the magnitude of the ratio exceeds approximately $ 5.2$, delocalizaion of the PSR is observed. And the OSR is localized in specific areas. The topological transition occurs from $C=-1$ to $C=2$ where ${\lambda_v}/{t_2} = 2.8$. Analogous to the QWZ model, the tendency of delocalization within OSR does not change between non-trivial phase. On the contrary, when the Chern number is non-trivial ($- 5.2 < \frac{\lambda_v}{t_2} < 5.2$), the OSR exhibit delocalization.

	\begin{figure}[H]
		\centering
		\includegraphics[width =  1\linewidth]{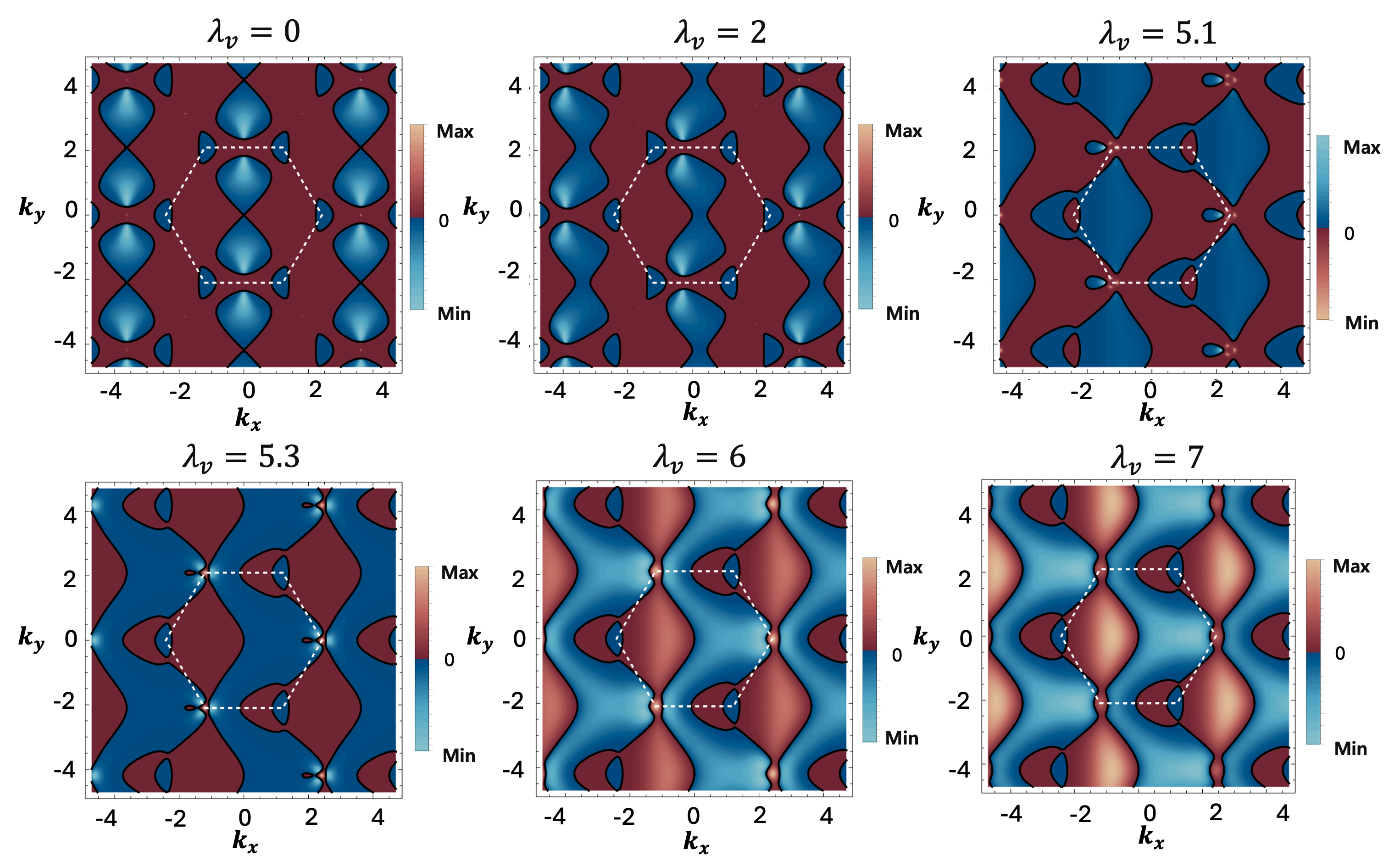}
		\caption{Berry curvature of the Extended Haldane model when $t_1 = 1, t_2= 1, t_3=0.35 , \phi =-\pi/2 $ : Dashed line represents the  BZ of the Extended Haldane model. The red region is peak signed region and the blue region is the oppositely signed reigion. When the $\lambda_v <5.2$, the topological phase is non trivial, therefore the delocalization of the red region occurs while the blue region is delocalized. On the other hand, the topological phase is trivial, the delocalization occurs within the blue regions like a river. With the topological transition, the percolation of the blue regions occurs.}
		\label{fig:N3Haldaneberrycurvature}
	\end{figure}

	\begin{figure}[H]
	\includegraphics[width =  0.6 \linewidth]{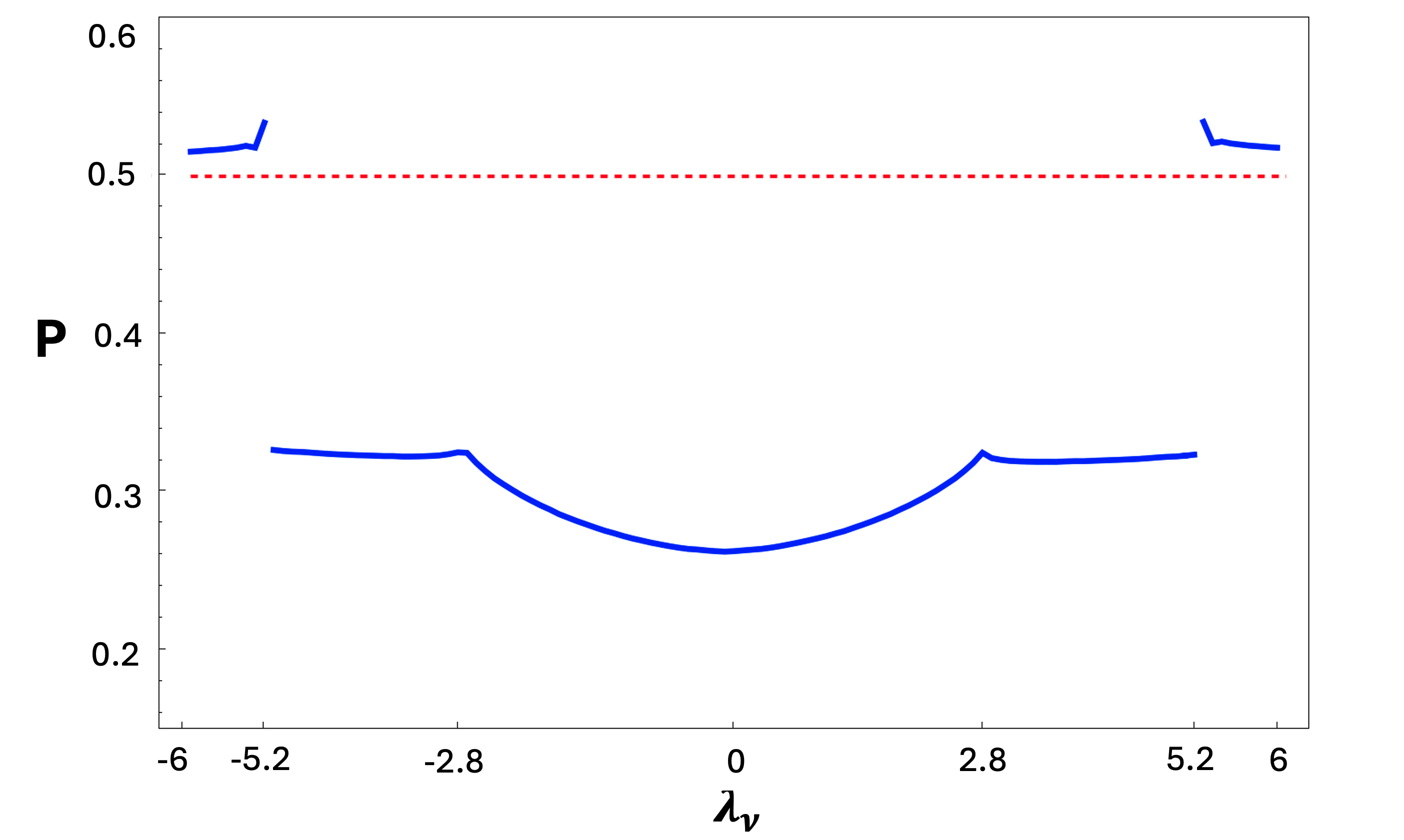}
	\centering
	\caption{Percolation rate of the OSR in the Extended Haldane model. When the $-5.2>\lambda_v \ or \  5.2 <\lambda_v$, the Chern number is trivial, the percolation rate is higher than 0.5 which means that delocalization occurs dominantly in the BZ. On the other hand, the Chern number is non trivial, the percolation rate is lower than the 0.5.}
	\label{fig:N3PR}
\end{figure}

		Based on the results, we calculate the percolation rate using Eq (\ref{percolationRate}). In Fig. \ref{fig:QWZPR}, the red line represents $P=0.5$, while the blue line indicates the percolation rate of the OSR in the QWZ model. When $C=0$, the percolation rate exceeds 0.5, indicating that the OSR are predominantly delocalized in the BZ. Conversely, when C is non zero, the percolation rate is below 0.5. Consequently, the distribution of the Berry curvature in both OSR and PSR acts as the topological feature that determines whether the topological phase is trivial or non-trivial.

	\section{Discussions}
	
	In this work, we have demonstrated that the percolation behavior of the Berry curvature can serve as a pivotal role in determining the topological phases of materials. By analyzing the Berry curvature distribution through the perspective of topological transitions, we have established a clear and practical criterion for distinguishing whether the Chern number is trivial or non-trivial. This approach was validated through comprehensive analyses of the QWZ model, Haldane model, and Extended Haldane model over extensive  parameter regime.
	
The key insight of our study is the introduction of relative sign analysis  by defining  Peak Signed Regions (PSR) and Oppositely Signed Regions (OSR). This approach reveals global patterns in the Berry curvature that align closely with topological phase transitions.
		We observed  the  delocalization  of OSR  over the Brillouin zone (BZ) when the  topology  is trivial. In contrast, for non-trivial  topology, PSR is delocalized  while  OSR is  localized. 
	The transition between trivial and non-trivial topological phases is marked by a clear change in the percolation characteristics of PSR and OSR. This observation provides a framework for predicting and understanding the topological properties of materials. Furthermore, the global behavior of the Berry curvature distribution, as captured by our approach, supports future experimental validation.
	
	Efforts to directly measure Berry curvature, such as those involving cold atoms and photonic lattices\cite{Coldatom,Exphotonic}, show potential in characterizing topological properties. These measurements provide data for developing materials with tailored electronic properties and deepen our understanding of quantum and topological physics. Observing the percolation behavior of Berry curvature may enable more precise determinations of Chern numbers and facilitate the characterization of topological phases in experimental setups.
	
	Our findings represent observational results rather than formally proven statements. Additionally, this study focuses on cases with a single pair of bands and a single non-degenerate dominant Berry peak within the Brillouin zone. Future work could extend these findings to multi-band systems and examine the robustness of percolation behavior under different experimental conditions.
	
	We hope that relative sign analysis will advance the study of topological phases. This study emphasizes the significance of Berry curvature, even before its integration over the BZ.
	\bibliography{SciPostPhys}

\end{document}